\documentstyle[12pt]{article}

\def\journal{\topmargin .3in	\oddsidemargin .5in
	\headheight 0pt	\headsep 0pt
	\textwidth 5.625in 
	\textheight 8.25in 
	\marginparwidth 1.5in
	\parindent 2em
	\parskip .5ex plus .1ex		\jot = 1.5ex}
\journal

\catcode`\@=11
\def\marginnote#1{}
\newcount\hour
\newcount\minute
\newtoks\amorpm
\hour=\time\divide\hour by60
\minute=\time{\multiply\hour by60 \global\advance\minute by-\hour}
\edef\standardtime{{\ifnum\hour<12 \global\amorpm={am}%
	\else\global\amorpm={pm}\advance\hour by-12 \fi
	\ifnum\hour=0 \hour=12 \fi
	\number\hour:\ifnum\minute<10 0\fi\number\minute\the\amorpm}}
\edef\militarytime{\number\hour:\ifnum\minute<10 0\fi\number\minute}
\def\draftlabel#1{{\@bsphack\if@filesw {\let\thepage\relax
   \xdef\@gtempa{\write\@auxout{\string
      \newlabel{#1}{{\@currentlabel}{\thepage}}}}}\@gtempa
   \if@nobreak \ifvmode\nobreak\fi\fi\fi\@esphack}
	\gdef\@eqnlabel{#1}}
\def\@eqnlabel{}
\def\@vacuum{}
\def\draftmarginnote#1{\marginpar{\raggedright\scriptsize\tt#1}}
\def\draft{\oddsidemargin -.5truein
	\def\@oddfoot{\sl preliminary draft \hfil
	\rm\thepage\hfil\sl\today\quad\militarytime}
	\let\@evenfoot\@oddfoot	\overfullrule 3pt
	\let\label=\draftlabel
	\let\marginnote=\draftmarginnote
   \def\@eqnnum{(\theequation)\rlap{\kern\marginparsep\tt\@eqnlabel}%
\global\let\@eqnlabel\@vacuum}  }
\def\preprint{\twocolumn\sloppy\flushbottom\parindent 2em
	\leftmargini 2em\leftmarginv .5em\leftmarginvi .5em
	\oddsidemargin -.5in	\evensidemargin -.5in
	\columnsep .4in	\footheight 0pt
	\textwidth 10in	\topmargin  -.4in
	\headheight 12pt \topskip .4in
	\textheight 7.1in \footskip 0pt
	\def\@oddhead{\thepage\hfil\addtocounter{page}{1}\thepage}
	\let\@evenhead\@oddhead	\def\@oddfoot{}	\def\@evenfoot{} }
\def\numberbysection{\@addtoreset{equation}{section}
	\def\theequation{\thesection.\arabic{equation}}}
\def\underline#1{\relax\ifmmode\@@underline#1\else
	$\@@underline{\hbox{#1}}$\relax\fi}
\def\titlepage{\@restonecolfalse\if@twocolumn\@restonecoltrue\onecolumn
     \else \newpage \fi \thispagestyle{empty}\c@page\z@
	\def\thefootnote{\fnsymbol{footnote}} }
\def\endtitlepage{\if@restonecol\twocolumn \else \newpage \fi
	\def\thefootnote{\arabic{footnote}}
	\setcounter{footnote}{0}}  
\catcode`@=12
\relax
\def\figcap{\section*{Figure Captions\markboth
	{FIGURECAPTIONS}{FIGURECAPTIONS}}\list
	{Figure \arabic{enumi}:\hfill}{\settowidth\labelwidth{Figure 999:}
	\leftmargin\labelwidth
	\advance\leftmargin\labelsep\usecounter{enumi}}}
 \relax
\def\tablecap{\section*{Table Captions\markboth
	{TABLECAPTIONS}{TABLECAPTIONS}}\list
	{Table \arabic{enumi}:\hfill}{\settowidth\labelwidth{Table 999:}
	\leftmargin\labelwidth
	\advance\leftmargin\labelsep\usecounter{enumi}}}
 \relax
\def\reflist{\section*{References\markboth
	{REFLIST}{REFLIST}}\list
	{[\arabic{enumi}]\hfill}{\settowidth\labelwidth{[999]}
	\leftmargin\labelwidth
	\advance\leftmargin\labelsep\usecounter{enumi}}}
 \relax
\makeatletter
\newcounter{pubctr}
\def\publist{\@ifnextchar[{\@publist}{\@@publist}}
\def\@publist[#1]{\list
	{[\arabic{pubctr}]\hfill}{\settowidth\labelwidth{[999]}
	\leftmargin\labelwidth
	\advance\leftmargin\labelsep
	\@nmbrlisttrue\def\@listctr{pubctr}
	\setcounter{pubctr}{#1}\addtocounter{pubctr}{-1}}}
\def\@@publist{\list
	{[\arabic{pubctr}]\hfill}{\settowidth\labelwidth{[999]}
	\leftmargin\labelwidth
	\advance\leftmargin\labelsep
	\@nmbrlisttrue\def\@listctr{pubctr}}}
 \relax
\makeatother
\catcode`\@=11
\def\section{\@startsection {section}{1}{0pt}{-3.5ex plus -1ex minus
 -.2ex}{2.3ex plus .2ex}{\raggedright\large\bf}}
\catcode`\@=12
\newskip\humongous \humongous=0pt plus 1000pt minus 1000pt
\def\caja{\mathsurround=0pt}

\newif\ifdtup
\def\panorama{\global\dtuptrue \openup1\jot \caja
	\everycr{\noalign{\ifdtup \global\dtupfalse
	\vskip-\lineskiplimit \vskip\normallineskiplimit
	\else \penalty\interdisplaylinepenalty \fi}}}
\def\eqalignno#1{\panorama \tabskip=\humongous
	\halign to\displaywidth{\hfil$\displaystyle{##}$
	\tabskip=0pt&$\displaystyle{{}##}$\hfil
	\tabskip=\humongous&\llap{$##$}\tabskip=0pt
	\crcr#1\crcr}}

\def\oldreffmt#1{\rlap{[#1]} \hbox to 2\parindent{}}

\def\figfmt#1{\rlap{Figure {#1}} \hbox to 1in{}}

\def\abs#1{\left| #1\right|}

\def\beq{\begin{equation}}
\def\eeq{\end{equation}}

\def\bea{\begin{eqnarray}}

\def\eea{\end{eqnarray}}
\catcode`\@=11
\def\eqnarray{\stepcounter{equation}\let\@currentlabel=\theequation
\global\@eqnswtrue
\global\@eqcnt\z@\tabskip\@centering\let\\=\@eqncr
\gdef\@@fix{}\def\eqno##1{\gdef\@@fix{##1}}%
$$\halign to \displaywidth\bgroup\@eqnsel\hskip\@centering
  $\displaystyle\tabskip\z@{##}$&\global\@eqcnt\@ne
  \hskip 2\arraycolsep \hfil${##}$\hfil
  &\global\@eqcnt\tw@ \hskip 2\arraycolsep $\displaystyle\tabskip\z@{##}$\hfil
   \tabskip\@centering&\llap{##}\tabskip\z@\cr}

\def\@@eqncr{\let\@tempa\relax
    \ifcase\@eqcnt \def\@tempa{& & &}\or \def\@tempa{& &}
      \else \def\@tempa{&}\fi
     \@tempa \if@eqnsw\@eqnnum\stepcounter{equation}\else\@@fix\gdef\@@fix{}\fi
     \global\@eqnswtrue\global\@eqcnt\z@\cr}

\catcode`\@=12
\font\tenbifull=cmmib10 
\font\tenbimed=cmmib10 scaled 800
\font\tenbismall=cmmib10 scaled 666
\def\boldzeta{\fam=9{\mathchar"7110 } }

\def\boldlambda{\fam=9{\mathchar"7115 } }

\relax

\def\thefootnote{\fnsymbol{footnote}}
\begin{document}
\begin{titlepage}
\begin{center}
\today     \hfill    LBL-36022 \\
          \hfill    UCB-PTH-94/22 \\

\vskip .25in

{\large \bf Signals For Supersymmetric Unification}
\footnote{This work was supported in part by the Director, Office of
Energy Research, Office of High Energy and Nuclear Physics, Division of
High Energy Physics of the U.S. Department of Energy under Contract
DE-AC03-76SF00098 and in part by the National Science Foundation under
grant PHY-90-21139.}
\vskip .25in
R. Barbieri\\

{\em Department of Physics\\
University of Pisa\\
and\\
INFN, Pisa, Italy.}

\vskip .25in
L.J. Hall

{\em Department of Physics\\
     University of California\\
    and\\
    Theoretical Physics Group\\
    Lawrence Berkeley Laboratory\\
    Berkeley, California 94720}
\end{center}

\vskip .25in

\begin{abstract}

Using the Yukawa couplings of the minimal supersymmetric SU(5) model, the rates
for $\mu \to e\gamma$, $\mu \to e$ conversion and $\tau \to \mu \gamma$
are computed.
For a selectron mass of 100 GeV, and without exploring the full parameter
space,
we find rates which are one
order of magnitude beneath present experimental bounds.
It is argued that these relatively large rates have a
wide applicability, so that lepton flavor violating signals provide a more
general
test of supersymmetric unification than can be obtained
from either proton decay or neutrino masses.

\end{abstract}
\end{titlepage}
\renewcommand{\thepage}{\roman{page}}
\setcounter{page}{2}
\mbox{ }

\vskip 1in

\begin{center}
{\bf Disclaimer}
\end{center}

\vskip .2in

\begin{scriptsize}
\begin{quotation}
This document was prepared as an account of work sponsored by the United
States Government. While this document is believed to contain correct
 information, neither the United States Government nor any agency
thereof, nor The Regents of the University of California, nor any of their
employees, makes any warranty, express or implied, or assumes any legal
liability or responsibility for the accuracy, completeness, or usefulness
of any information, apparatus, product, or process disclosed, or represents
that its use would not infringe privately owned rights.  Reference herein
to any specific commercial products process, or service by its trade name,
trademark, manufacturer, or otherwise, does not necessarily constitute or
imply its endorsement, recommendation, or favoring by the United States
Government or any agency thereof, or The Regents of the University of
California.  The views and opinions of authors expressed herein do not
necessarily state or reflect those of the United States Government or any
agency thereof or The Regents of the University of California and shall
not be used for advertising or product endorsement purposes.
\end{quotation}
\end{scriptsize}

\vskip 2in

\begin{center}
\begin{small}
{\it Lawrence Berkeley Laboratory is an equal opportunity employer.}
\end{small}
\end{center}

\newpage
\renewcommand{\thepage}{\arabic{page}}
\setcounter{page}{1}

{\bf 1.}

Grand unified theories [1] provide an elegant unification of the strong
and electroweak forces, and, with the addition of supersymmetry, lead to a
successful prediction of the weak mixing angle [2] at the 1\% level of
accuracy.
While weak scale supersymmetry can be directly probed by the production of
superparticles in high energy collisions, signals for grand unification are
less
direct.

Experiments probe grand unified theories  either by
relating parameters of the standard model which are otherwise independent or
by producing additional interactions which mediate processes which are
forbidden, or highly suppressed, in the standard model.
The weak mixing angle, and the $m_b/m_\tau$ mass ratio are examples of the
former, while proton decay is an example of the latter.
In this letter we concentrate on the rare process signals of supersymmetric
grand unification.
We claim that lepton flavor-violating processes provide
the signatures which can be most reliably calculated and which are
most generic to the idea of grand unification.
Using the Yukawa couplings of
the minimal supersymmetric $SU(5)$ theory
and a selectron mass of 100 GeV, we find rates for these processes which are
one
order of magnitude beneath present experimental bounds.

In the standard model, baryon number ($B$), the individual lepton numbers
($L_e,
L_\mu, L_\tau$) and the total lepton number $(L = L_e + L_\mu + L_\tau$) are
accidental symmetries and therefore, apart from very small instanton effects,
are conserved. On the other hand, grand unification removes any fundamental
distinction between a quark and a lepton, so that at the unification scale,
$M_G$, these symmetries are broken.
We introduce three categories of signatures,
and label them according to the symmetries they violate

$B$: \hskip 9pt  proton decay, etc.

$L$:  \hskip 9pt neutrino masses, etc.

$L_i$: \hskip 10pt $\mu \to e\gamma, $ etc.

If the effective theory beneath $M_G$ is the standard model, the phenomena of
all three categories are suppressed by powers of $1/M_G$.
This does not exclude the possibility of experimentally probing
$B$ and $L$ violation, since the
experimental sensitivity to proton decay and to neutrino masses is such that
even effects power suppressed by $1/M_G$ may be observable.
However, for $L_i$ violating processes, such as $\mu \to e\gamma$, present
experiments are many orders of magnitude away from having sensitivity to
effects
suppressed by powers of $1/M_G$.
Thus the conventional signatures for grand unification have been taken to be
those which violate $B$ or $L$, for example proton decay and neutrino masses.

If the effective theory beneath $M_G$ is supersymmetric, for example the
minimal
supersymmetric standard model (MSSM), a similar line of reasoning leads to a
very different conclusion. The most general gauge invariant low energy theory
now contains renormalizable interactions which violate
$B, L, \  \hbox{and}\  L_i$.
Such interactions induce proton decay at an unacceptable rate, so the gauge
symmetries are augmented by a further symmetry, called $R$ parity or matter
parity, which forces all interactions to have an even number of quarks and
leptons and their superpartners.
Consequently, the renormalizable interactions beneath $M_G$ conserve $L$ and
$B$, but violate $L_i$.
Thus in many supersymmetric theories phenomena which violate $B$ or $L$ are
suppressed by powers of $1/M_G$, while processes which violate $L_i$ are
suppressed only by powers of $1/m$, where $m$ is the scale of supersymmetry
breaking.
Since $m \approx 100$ GeV while $M_G \approx 10^{16}$ GeV, the $L_i$ violation
processes now become a competitive experimental probe.

The operators which violate $L_i$ are slepton masses and trilinear scalar
interactions, and have the general form:

$$
V_{\triangle L_i} = E^+{\bf m}^2_E E + L {\bf m}^2_L L^+
+  (L {\boldzeta}_E E H_1 + h.c.)  \eqno(1)
$$
where $L$ and $E$ are 3-vectors containing the $SU(2)$ doublet and singlet
sleptons, taken in a basis where the leptons are mass eigenstates.
$H_1$ is a Higgs doublet. The parameters
${\bf m}^2_E, {\bf m}^2_L$ and ${\boldzeta}_E$ are $3\times 3$ matrices.
The size of the off diagonal entries of the matrices now becomes the crucial
question.
It is well known that these quantities must be restricted in size: off-diagonal
entries in ${\bf m}^2_{E,L}$ of order of the square of the average slepton mass
are experimentally excluded.
In the MSSM, based on supergravity with supersymmetry broken in a hidden
sector [3],
it is assumed that at the Planck scale a boundary condition on the
theory forces ${\bf m}^2_E, {\bf m}^2_L$ to be
proportional to the unit matrix,  and ${\boldzeta}_E = A {\boldlambda}_E$ where
${\boldlambda}_E$ is the lepton Yukawa matrix. \footnote{We will assume such
boundary conditions throughout this paper.}
In this case the theory conserves $L_i$ exactly.
However, if the theory becomes grand unified above $M_G$, it is not possible to
have $L_i$ conservation.
This is because flavor changing processes occur in the quark sector and grand
unification connects the quark and lepton sectors thus introducing $L_i$
violation. Furthermore, the flavor violations
of the grand unified interactions do not decouple at low energies,
as first
shown in reference 4,
but manifest
themselves via soft operators such as those shown in equation (1).

How large are the $L_i$ violating interactions ${\bf m}^2_{E,L}$ and
${\boldzeta}_E$ generated by supersymmetric unified theories?
In this letter we compute precise formulae  for the $L_i$ violating effects
using the Yukawa couplings of
the minimal supersymmetric $SU(5)$ theory.
We discuss the uncertainties in the calculation and explain why we believe our
calculation represents a lower bound on what is to be expected from a generic
supersymmetric grand unified theory.
While it is possible to construct theories in which the $L_i$ violation is
suppressed, a large suppression requires some special ad hoc arrangement, and
is
not the general expectation.
In fact, it is very much easier to write down theories where the $L_i$
violating
effects are much larger than those of the minimal model.
The uncertainties in our result are not large; not nearly
as large or model dependent as those
involved in calculating either the proton decay rate or neutrino masses.

The reasons for this will be discussed later in some detail.
However, one reason is that $B$ and $L$ violating processes depend on a power
of
$1/M_G$, and the relevant value for $M_G$ which should appear in the rate for
these processes is not well known.
On the other hand, the $L_i$ violating processes do not have any power
dependence on $1/M_G$, and hence are much less sensitive to this mass scale.
We believe that the rare processes from supersymmetric unification which can be
calculated most reliably, and which are most generic, are $L_i$ violating
processes such as $\mu \to e \gamma$.

The mechanism [4] which
we use to compute the $L_i$ violating matrices of (1)
gives a large result only because the top Yukawa coupling is large.
In the minimal $SU(5)$ theory the up and down quark masses are generated by
$T{\boldlambda}_U TH$ and $\overline{F} {\boldlambda}_D T\overline{H}$
where $T$ and
$\overline{F}$ are three vectors of quarks and leptons and their superpartners
in the 10 and $\overline{5}$ dimensional representations of $SU(5)$, and $H$
and
$\overline{H}$ are 5 and $\overline{5}$ representations of Higgs multiplets.
\footnote{The decomposition of
$T, \overline{F}, H$ and $\overline{H}$ are $T\supset Q, U,
E; \ \overline{F} \supset\overline{D}, L; \ H \supset H_2 H_3$;
$\overline{H} \supset H_1,
\overline{H}_3$ where $Q, L$ are $SU(2)$ doublet quarks and leptons, $U, D, E$
are $SU(2)$ singlet quarks and leptons, $H_1$ and $H_2$ are the two $SU(2)$
Higgs doublets of the MSSM and their $SU(5)$ partners are the color triplet
fields $\overline{H}_3$ and $H_3$.}
The only flavor violation of the model comes from the $3\times 3$ matrices
${\boldlambda}_U$ and ${\boldlambda}_D$.
The matrix ${\boldlambda}_U$ has a large eigenvalue which is responsible for
the top
quark mass. The interaction $T {\boldlambda}_U TH$ contains not only couplings
to the Higgs doublet $H_2$ in $H$ which are responsible for masses:
$U{\boldlambda}_U Q H_2$, but also the couplings to a Higgs triplet $H_3$
$$
\Delta W     = U {\boldlambda}_U E H_3.\eqno(2)
$$
This new interaction is a necessary consequence of unifying the quarks and
leptons in $SU(5)$ representations, and furthermore, the $SU(5)$ symmetry
requires that the matrix of couplings, ${\boldlambda}_U$, is identical to
those which appear in the mass coupling $U {\boldlambda}_U Q H_2$.
Thus the theory possesses $L_i$ violation via the large top Yukawa coupling.

Previous considerations [4,5] of $L_i$ violating effects using the interaction
(2) have claimed values too small to be of interest.
In this letter we find that the contribution from the top Yukawa
coupling is in fact
of great interest.

{\bf 2.}

The scalar mass parameters, {\bf{m}}$^2_{L, E}$, and the scalar trilinear
couplings, ${\boldzeta}_E$, depend on energy scale, as given by the
renormalization group equations.
In the MSSM these equations preserve $L_i$ conservation.
However, when the interaction of eq. (2) is added to the MSSM the
renormalization group equations of the grand unified theory, valid from energy
scales of the Planck scale, $M_{Pl}$, down to $M_G$, generate $L_i$ violating
masses:
$$
\triangle m^2_{E_{ij}} =
-  V^*_{ti}V_{t_j} \; I \eqno(3)
$$
where the integral I is
$$
I = {3 \over 8 \pi^2} \int \lambda_t^2
(m_H^2 + 2 m_{T_3}^2 + A_t^2) d(ln \mu).  \eqno(4)
$$
This integral is evaluated from $M_G$ to $M_{Pl}$, $m_H$ is the mass of the
scalars in H, $m_{T_3}$ is the mass of the scalars in $T_3$ and the coefficient
of the scalar trilinear interaction $T_3 T_3 H$ is $A_t \lambda_t$. The Yukawa
coupling $\lambda_t$ is the large eigenvalue of the matrix  ${\boldlambda}_U$
and, both here and elsewhere, we drop smaller contributions from
${\boldlambda}_U$ and  ${\boldlambda}_D$.
{\bf V} is the Kobayashi-Maskawa matrix, renormalized at $M_G$.
The solution of the renormalization group equation for the slepton trilinear
interaction can be written in a similar form if the scaling of {\bf V} is
ignored
$$
\triangle \zeta_{E_{ij}} =  V^*_{ti}V_{t_j} \; I_i' \eqno(5)
$$
where
$$
I_i' = {3 \over 8 \pi^2} \; \int A_t \lambda_t^2 \lambda_{e_i} d(ln \mu)
\eqno(6)
$$
and $\lambda_{e_i}$ is the i th eigenvalue of ${\boldlambda}_E$.

To illustrate how large the flavor changing slepton masses from the top Yukawa
coupling can be, we study the renormalization group equations in the limit that
the contributions to scalar masses from gaugino masses can be neglected. In
this case we find
$$
I = m^2_0 \; (1 - e^{-J}) \eqno(7)
$$
where
$$
J = {3 \over 8 \pi^2} \int \lambda_t^2 (3 + {\abs{A_t}^2 \over m^2}) d(ln \mu)
 \eqno(8)
$$
where $m_{T_3} = m_H =m$, and the value of m at $M_{Pl}$ is $m_0$.
If $\lambda_t$ is small then J is small, so that the
exponent of equation (7) can be expanded to give the one-loop result
$$
I \simeq  {3 \over 8 \pi^2}  \lambda^2_{tG} (3 m_G^2 + \abs{A_{t_G}}^2)
 \ln {M_{Pl} \over M_G}\eqno(9)
$$
where $m_G^2, \lambda_{t_G}$ and $A_{t_G}$
are the values of $m^2,  \lambda_t$ and $A_t$ at $M_G$.
In the case of a small gaugino mass the physical masses of the right-handed
sleptons are given by  $m_{\tilde{\tau}}^2 =  m_G^2$ and
$m_{\tilde{e}}^2 =  m_{\tilde{\mu}}^2 =  m_0^2$. Thus the right-handed scalar
tau is the lightest slepton, with
$$
{m_{\tilde{\tau}}^2 \over m_{\tilde{\mu}}^2} =  {m_G^2 \over  m_0^2}
= e^{-J} \eqno(10)
$$

One of the generic features of many grand unified theories is the $m_b/m_\tau$
prediction [6] which is known to be acceptable, for non-extreme values of
$\tan \beta$, only for
large values of $\lambda_{tG}$ [7], greater than approximately 1.5.
This allows us to derive a lower bound on J, and also on I. We assume
$\lambda_t \geq 1.5$ over the interval from $M_G$ to $M_{Pl}$. This gives $J
\geq 1.2$ and $I \geq 0.7 m_0^2$. Our choice of $\lambda_{t_G}$ is
the lowest allowed by
the  $m_b/m_\tau$ prediction, and in the minimal model $\lambda_t$ grows above
$M_G$. Furthermore the bound is obtained by setting A to zero, and is clearly
conservative. Nevertheless as J increases above this bound, I can only increase
to its saturation value of $m_0^2$, so that the predictions for flavor changing
effects have a great insensitivity to the details of the scaling above $M_G$.
For numerical estimates we will use $I= m_0^2$, since this is our expectation
for the majority of models.
The  $m_b/m_\tau$ prediction is crucial to our belief that the $L_i$ violating
signatures are likely to be seen in the next generation of experiments. If
$\lambda_{tG} = 0.4$, which is consistent with $m_t = 174$ GeV,
the signals would probably
be too small to be observed, but in this
case the  $m_b/m_\tau$ prediction fails.

Finally we note that $\triangle {\bf m}^2_L$ receives no large $\lambda^2_{tG}$
correction in the minimal model.

At the weak scale, the flavor violating interactions
$\triangle {\bf{ m}}^2_E$ and $\triangle{\boldzeta}_E$ can
be inserted into one loop diagrams involving internal superpartners to
yielding a Feynman amplitude for $\mu \to e \gamma^*$
$$
A^\mu = -i { e \over  m_{\tilde{\mu}}^2} \overline{u}_e \left( q^2
F'_1 \gamma^\mu P_R + m_\mu F_2 i\sigma^{\mu \nu} q_\nu P_L\right)
u_\mu\eqno(11)
$$
where $q_\nu$ is the 4-momentum of $\gamma^*$, $m_\mu$ the muon mass and
$P_{L,R}$ the chirality projection
 operators.
After calculating the relevant 1-loop diagrams we find the form factors to be
given, in the zero gaugino mass limit, by
$$
F'_1 = { \alpha (M_Z)\over 4\pi \cos^2\theta_W}\; {1\over 9} \;
 {\triangle m^2_{E_{21}}\over   m_{\tilde{\mu}}^2}\eqno(12)
$$
and
$$
F_2 = {\alpha (M_Z)\over 4\pi\cos^2\theta_W}  \;
 {1 \over 6} {\triangle m^2_{E_{21}}\over   m_{\tilde{\mu}}^2} \eqno(13)
$$
For the case of non-zero gaugino mass, we consider the example of
the bino a mass
eigenstate and degenerate with the selectron.
We find that in
equation (12) the factor of ${1 \over 9}$ is replaced by
 ${1 \over 15}$, while in equation (13) the
factor of ${1 \over 6} {\triangle m^2_{E_{21}}\over   m_{\tilde{\mu}}^2}$
is replaced by
$$
 \left({1 + {A \over  m_{\tilde{\mu}}} \over 20}
{\triangle m^2_{E_{21}}\over   m_{\tilde{\mu}}^2}
+ {1\over 12}
{\triangle {\zeta}_{E_{21}} \over  m_{\tilde{\mu}}}
{v_1\over m_\mu}\right)\eqno(14)
$$
where $v_1$ is the vev of $H_1$, and renormalization effects from $M_G$ to
$ m_{\tilde{\mu}}$ have not been included.

For the decay $\mu \to e\gamma$ the photon must be taken on shell so
only $F_2$ contributes
$$
\Gamma (\mu \to e \gamma) = {\alpha\over 4} \abs{F_2}^2 {m_\mu^5 \over
m_{\tilde{\mu}}^4}
. \eqno(15)
$$

For $\mu \to e$ conversion the Feynman amplitude from Penguin-like diagrams is,
in the zero gaugino mass limit
\newpage
$$
\eqalignno{
A(\mu \to e)_{Penguin} &= i  { e^2 \alpha (M_Z)\over 4\pi \cos^2\theta_W}\;
  {\triangle m^2_{E_{21}}\over   m_{\tilde{\mu}}^4} \cr
&\cdot \overline{u}_e ( {1 \over 9} \gamma^\nu P_R -
{1 \over 6} i \sigma^{\nu\mu} q_{\mu} P_L)
u_\mu \;
({2\over 3} \overline{u}_u \gamma_\nu u_u - {1\over 3}
\overline{u}_d \gamma_\nu u_d).&(16a)\cr}
$$

The $\mu \to e $ conversion amplitude receives also a contribution from box
diagrams, which is found to be, for a squark mass $ m_{\tilde{q}}^2 \gg
m_{\tilde{\mu}}^2$,
$$
A(\mu \to e)_{Box} = i  { e^2 \alpha (M_Z)\over 4\pi \cos^4\theta_W}\;
  {\triangle m^2_{E_{21}}\over   m_{\tilde{\mu}}^2   m_{\tilde{q}}^2 }\;
\overline{u}_e  \gamma^\nu P_R  u_\mu \;
({17\over 72} \overline{u}_u \gamma_\nu u_u + {5\over 72}
\overline{u}_d \gamma_\nu u_d).\eqno(16b)
$$
where only the vector piece of the quark current is kept. From the sum of both
amplitudes one finds a $\mu \to e $ conversion rate in $ Ti^{48}_{22}$ of
$$
\eqalignno{
\Gamma(\mu \to e; Ti) &= 4 \alpha^5 Z^4_{eff} ZF(q)^2
\left( { \alpha (M_Z)\over 4\pi \cos^2\theta_W} \;
 {\triangle m^2_{E_{21}}\over   m_{\tilde{\mu}}^2}\right)^2 \;
 {m_\mu^5\over m^4} \cr
&\cdot \abs{{1 \over 18} +{ 39 + 27 {N \over Z} \over 72 \cos^2 \theta_W}
{  m_{\tilde{\mu}}^2 \over   m_{\tilde{q}}^2} }^2
&(17)\cr}
$$
where $Z_{eff} = 17.6, Z=22, N = 26$ and $F(q) = .54 [8]$.

Evaluating the rates (15) and (17), using (13) for $F_2$ and
 $\triangle m^2_{E_{21}}$
from eqs. (3,7), gives our results in the limiting case of vanishing gaugino
mass:
$$
BR(\mu\to e\gamma) = 2.4 \times 10^{-12}
   \left( { \abs{V_{ts}}\over 0.04}
 { \abs{V_{td}}\over 0.01}\right)^2
\left( {100 GeV\over m_{\tilde{\mu}}}\right)^4
\eqno(18)
$$
and for the $\mu \to e $ conversion rate, normalized to the capture rate [8],
in $Ti$:
$$
R(\mu \to e; Ti) = 5.1 \times 10^{-14}
 \left( {\abs{V_{ts}}\over 0.04} {\abs{V_{td}}\over 0.01}\right)^2
 \left( {100 GeV\over  m_{\tilde{\mu}}}\right)^4 \eqno(19)
$$
where we have taken  $m_{\tilde{q}} = 2 m_{\tilde{\mu}}$. We ignore the
renormalization group scaling of ${\bf V}$ from $M_G$ to $ m_{\tilde{\mu}}$,
so in
these results $V_{ij}$ are the measured values of the Kobayashi-Maskawa mixing
matrix.
The formula for $\tau \to \mu \gamma$ follows directly
from similar equations to those above
$$
BR(\tau \to \mu\gamma) = {\abs{V_{tb}}^2\over\abs{V_{td}}^2}\;
\left({  m_{\tilde{\mu}} \over   m_{\tilde{\tau}} }\right)^4
BR(\mu \to e\gamma),\eqno(20)
$$
giving
$$
BR(\tau \to \mu\gamma) = 3.8 \times 10^{-7}
 \left( {\abs{V_{ts}}\over 0.04} \right)^2
\left( {50 GeV\over  m_{\tilde{\tau}}}\right)^4, \eqno(21)
$$
where the choice for the normalization of the right-handed scalar tau mass is
motivated by eq. (10).
For the scalar masses we have chosen, $\mu \to e\gamma$ and $\mu\to e$
conversion
in $Ti$ are both a factor of 20 below the current experimental limits of
$4.9 \times 10^{-11}$ [9] and
$1 \times 10^{-12}$ [10] respectively, and
$\tau \to \mu \gamma$ a factor of 10 below the current experimental limit
of $4.2 \times 10^{-6}$ [11].

We have clearly not made an exploration of the full parameter space, but we
have rather concentrated on the simplifying case of a negligably small gaugino
mass. Inclusion of gaugino mass effects will bring in other parameters of the
low energy theory.
We stress, however, that  the majority of the uncertainties are ones which can
eventually be removed by measuring parameters at the weak
scale, and are not due to inherent limitations of the theory.
For a top mass of 174 GeV, the Yukawa coupling $\lambda_{t_G}$ must be greater
than 0.4. If  $\lambda_{t_G}$ were less than unity, the $L_i$ violating rates
would be smaller than given above, but in this case the $m_b/m_\tau$ prediction
fails.

If $\tan\beta$ is found to be large, of order $m_t/m_b$,
so that $\lambda_b$ and $\lambda_t$ are comparable, there will be extra
contributions to $\triangle{\bf{ m}}^2_E$ and $\triangle{\boldzeta}_E$
of order $\lambda^2_{bG}$ to added to equations (4) and (6).
These contributions would be comparable to the ones we have computed.
However, in the minimal $SU(5)$ theory $\triangle{\bf{ m}}^2_L$ remains zero.

{\bf 3.}

The flavor interactions of the minimal $SU(5)$ theory are known to be
incorrect as they yield the mass relation $m_e/m_\mu = m_d/m_s$.
Hence, a crucial question is whether the predictions of eqs. (18), (19),
and (21) survive in a general supersymmetric unified theory.
Any theory which contains the interaction of eq. (2) will yield these
predictions.
We know that the low energy theory must contain a Yukawa interaction
$U{\boldlambda}_U Q H_2$ to give mass to the up type quarks.
The grand unified theory must have interactions which incorporate this Yukawa
coupling.
The grand unified symmetry will require that there are other particles in the
representations which contain $U, Q, H_2$
and hence it is quite unavoidable that the grand unified interactions which
are responsible for $U{\boldlambda}_U Q H_2$ will lead to other interactions
also.
If $Q$ and $E$ are unified in the same representation, then an interaction
involving $UE H_3$ will necessarily occur.
Hence the most basic assumption is that quarks and leptons are unified in the
same representation.
Although it is possible to construct models in which $Q$ and $E$ lie in
different representations of the grand unified theory,
these models are artificial and completely destroy the aesthetic
appeal of quark-lepton unification.

The next question is whether the structure of the grand unified theory could
be such that the interactions which lead to $UEH_3$
vertices somehow conserve generation number.
This is possible. Consider,
for example, an $SU(5)$ model in which the $H$ representation becomes a matrix
of fields in generation space, so that the relevant couplings are
$T_i {\boldlambda}_{Uij} H_{ij} T_j$.
In this case the grand unified theory may have a symmetry which prevents any
violation of flavor.
However, at some scale, $M_F$, these flavor symmetries must be spontaneously
broken so that a non-trivial Kobayashi-Maskawa matrix, {\bf{V}},
results.
If $M_F \geq M_G$ our results survive, with $M_{Pl}$ in equation (4) and (6)
replaced by $M_F$.
However, if $M_F < M_G$, then flavor violating effects from the grand unified
theory must be suppressed by powers of $M_F/M_G$.
\footnote{In this case one expects
{\bf{m}}$^2_E$, {\bf{m}}$^2_L$ and ${\boldzeta}_E$ to
be generated at scale
$M_F$ so that lepton flavor violating processes are expected, but are
originated by physics at $M_F$.}
However, such a scheme would require several Higgs doublets at scale $M_F $
giving a prediction for the weak mixing angle in gross violation with data.

We conclude that in all grand unified theories, which have quarks and
leptons unified in the same representations and a successful
weak mixing angle prediction, our mechanism for $L_i$ violation is
necessarily present.
To what extent do they yield the precise formulas of equations (3,4) and (5,6)?
This depends on the form of the grand unified interactions.
Over some of the interval between $M_{Pl}$ and $M_G$, the interactions which
violate flavor could be linear
in quark and lepton fields, the usual bilinear operators only appearing after
integrating out heavy states.
In this case the $L_i$ violation is likely to be very much larger than
in the minimal model, but is very model dependent and we do not
consider it [12].
Over some of the interval between $M_{Pl}$ and $M_G$ we assume
that the interactions which violate flavor are  bilinear in quark and
lepton fields.
In any such theory we can write the relevant pieces of the interaction
in $SU(5)$ form as
$$
T {\boldlambda}_U \left({\Sigma \over M}\right) T H + \overline{F}
{\boldlambda}_D \left({\Sigma \over M}\right) T \overline{H}\eqno(22)
$$
where the Yukawa matrices are now functions of other fields $\Sigma$,
which acquire vevs of order $M_G$, giving fermion masses
suppressed by factors of $M_G/M$.
These operators have the virtue that, unlike those of minimal $SU(5)$, they
can lead to acceptable fermion masses.
Specific examples of such theories have been constructed [12] but here we
wish to study them in general.
The crucial point is that the fields $\Sigma$, if non-trivial under $SU(5)$,
lead to relative Clebsch factors between the couplings of the doublet $H_2$
and those of the triplet $H_3$.
Thus the matrix ${\boldlambda}_U$ in equation (2) must be replaced
with ${\boldlambda}'_U$ where the relation between the $ij$
entry of ${\boldlambda}_U$ and ${\boldlambda}'_U$ involves unknown Clebsch
factors.
In fact the 33 entries are still identical and equal to $\lambda_{tG}$:
the top Yukawa coupling is so large that it must come from a renormalizable
interaction and should not be suppressed by powers of $M_G/M$.
Thus the contribution from interaction (2) can be written as in equations
(3) and (4)
with $\lambda_{tG}$ unchanged, but with a mixing
matrix {\bf{V}}$'$ which is different from {\bf{V}}.
The relation between the $V'_{ij}$ and $V_{ij}$ involves the unknown Clebsch
factors.
Similarly the interaction (2) leads to a rate for the processes
$\mu \to e \gamma, \mu \to e$ conversion in $Ti$ and $\tau \to \mu
\gamma$ given by equations (18), (19), and (21), with $V_{ij} \to V'_{ij}$.
This leads to an increased range of the predictions about the central
values.
\footnote{Interactions of the form (22) also imply that the matrices
which couple $L$ to $H_1$ and $\overline{H}_3$ are
no longer identical: $L{\boldlambda}_E E H_1 + L
{\boldlambda}'_E Q\overline{H}_3$.
This allows a flavor changing $\triangle ${\bf{m}}$^2_L$ to be generated
proportional to ${\boldlambda}'^+_E {\boldlambda}'_E$.
For moderate $\tan\beta$ these couplings are much less than
 $\lambda_{tG}$ and can be neglected.
However, for large $\tan\beta \; (\approx m_t/m_b), \; m^2_{L_{21}}$ will
 become comparable to $m^2_{E_{21}}$.
This allows the $\mu \to e\gamma^*$ process to occur via a 1-loop diagram
involving a wino rather than a bino, so that in the rates
${\alpha^2\over \cos^4\theta}$ is replaced by ${\alpha^2\over \sin^4\theta}$,
 enhancing the
results by an order of magnitude.}

{\bf 4.}

Although our predictions have several uncertainties, these are far fewer
and less severe than those associated with the calculation of proton
decay.
Proton decay is mediated by the exchange of superheavy color triplets such as
$H_3$ and $\overline{H}_3$.
We list six types of uncertainty which enter the calculation of
$\Gamma (p\to K \nu)$ in supersymmetric unified theories

1) The mass scale of the color triplets.

2) The structure of the mass matrix for the superheavy triplets.

3) The mass matrix for the superheavy doublets.

4) The QCD matrix element.

5) The Clebsch factors arising from the couplings  of quarks and leptons to
\indent the Higgs triplets from operators such those  in equation (22).

6) Lack of knowledge of weak scale parameters, such as $V_{ij}, \tan\beta,
m, A,$ etc.

Of these six uncertainties, we believe the first three to be extremely
problematic.
While they can be largely overcome in the minimal $SU(5)$ theory,
they completely destroy the ability to make a prediction when certain
very minor additions to the theory are made, or in the case of a generic
unified theory.
Furthermore these three uncertainties are intimately connected to the
puzzle as to why the triplets $(H_3, \overline{H}_3 ...)$
are much heavier than the two light doublets $(H_1, H_2)$.
In the minimal $SU(5)$ theory this is accomplished by a fine tune which
greatly reduces the believability of the theory.
In extensions which solve this puzzle the first three uncertainties
generically destroy the predictability of the proton decay
rate [13].
In the minimal $SU(5)$ theory the central value of the prediction for
the proton decay rate is already clearly excluded, it is
therefore probably that some suppression mechanism must be operative.
In this case it is hard to argue how much suppression is to be expected.
Certainly it is straightforward to construct unified theories with proton decay
unobservably small [14].
Similarly, there are many $SU(5)$ and $SO(10)$ models where the light
neutrinos are exactly massless.

For the $L_i$ violating mechanism discussed here the first four uncertainties
listed above are absent.
We have discussed the remaining two uncertainties  at length.
If supersymmetry is discovered then the sixth uncertainty will eventually
be controlled, leaving only the unknown Clebsch factors.
This is the only real uncertainty which is inherent to the generic grand
unified theory.
However, the flavor operators of equation (22) generate both the quark and
lepton masses and mixings and the $L_i$ signals discussed here.
One might hope that eventually there will be enough experimental data
to determine a preferred set of operators.

Based on these considerations, we conclude that the lepton flavor violating
signals studied in this paper provide a very significant test of supersymmetric
unification. Such a test must be considered more general than can be obtained
from either proton decay or neutrino masses. At this point we would very much
like to know the ultimate experimental sensitivities that can be reached
in the processes that we have suggested. The present literature on the subject
is already encouraging [15].

\noindent{\bf Acknowledgements}

RB thanks the Miller Institute and the LBL theory group for hospitality and
partial support.

\noindent{\bf References}
\begin{enumerate}
\item H. Georgi and S. Glashow, {\it Phys. Rev. Lett.} {\bf{ 32}} (1974) 438.
\item S. Dimopoulos, S. Raby and F. Wilczek,{\it Phys. Rev.} {\bf{D 24}}
(1981) 1681; S. Dimopoulos and H. Georgi,{\it Nucl. Phys.} {\bf{ B193}} (1981)
150; L. Ibanez and G. G. Ross, {\it Phys. Lett.} {\bf{ B105}} (1981) 439.
\item R. Barbieri, S. Ferrara and C. Savoy,  {\it Phys. Lett.} {\bf{ B110}}
(1982) 343; P. Nath, R. Arnowitt and A. Chamseddine,  {\it Phys. Rev. Lett.}
{\bf{ 49}} (1982) 970; L.J. Hall, J. Lykken and S. Weinberg, {\it Phys. Rev.}
{\bf D 27} (1983) 2359.
\item L.J. Hall, V.A. Kostelecky and S. Raby,
{\it Nucl. Phys.} {\bf{ B267}} (1986) 415.
\item F. Gabbiani and A. Masiero, {\it Phys. Lett.} {\bf{ B209}} (1988) 289;
J.S. Hagelin, S. Kelley and T. Tanaka, {\it Nucl. Phys.}
{\bf{ B415}} (1994) 293.
\item M. Chanowitz, J. Ellis, and M.K. Gaillard, {\it Nucl. Phys.}
{\bf{ B128}} (1977) 506; A. Buras, J. Ellis, M.K. Gaillard, and D.
V. Nanopoulos, {\it Nucl. Phys. } {\bf{ B135}} (1978) 66.
\item L.E. Ibanez and C. Lopez, {\it Phys. Lett.} {\bf{ B126}} (1983) 54;
{\it Nucl. Phys.} {\bf{ B233}} (1984) 511; H. Arason et al.,
{\it Phys. Rev. Lett.} {\bf{ 67}} (1991) 2933; L.J. Hall and U. Sarid,
{\it Phys. Lett.} {\bf{ B271}} (1991) 138; S. Kelley, J. L.
Lopez and D.V. Nanopoulos, {\it Phys. Lett.} {\bf{ B274}} (1992) 387.
\item J. Bernab\'eu, E. Nardi, and D. Tommasini, {\it Nucl. Phys. } {\bf{
B409}}
(1993) 69, and references therein.
\item R. Bolton et al., {\it Phys. Rev.} {\bf D38} (1988) 2077.
\item SINDRUM II Collaboration, as quoted by R. Patterson, talk given at the
International Conference on High Energy Physics, Glasgow, July 1994.
\item A. Bean et al.,{\it Phys. Rev. Lett.} {\bf 70} (1993) 138.
\item G. Anderson, S. Dimopoulos, L.J. Hall, S. Raby and G. Starkman,
{\it Phys. Rev.} {\bf D49}  (1994) 3660; R. Barbieri, G. Dvali, A. Strumia, Z.
Berezhiani and L. Hall, LBL preprint - 35720, to appear in {\it Nucl. Phys.}
{\bf B}.
\item To recover a prediction, assumptions must be made which are sufficient
to determine the mass matrices for the triplets and
doublets.
One such set of assumptions is being studied in $SO(10)$ theories containing
the Dimopoulos-Wilczek mechanism for doublet-triplet splitting:
A. Antarian, L. J. Hall, H. Murayama and A. Rasin, preprint in preparation, LBL
35365.
\item K. Babu and S. Barr, {\it Phys. Rev.} {\bf D48} (1993) 5354.

\item J. F. Amann et al., Mega Experiment, preprint VPI-IHEP-93/7 (1993);
SINDRUM II Collaboration, preprint PSI-PR-90-41 (1990); MELC Collaboration, V.
Abadjev et al., preprint MOSCOW (1992).

\end{enumerate}

\end{document}